\documentclass[fleqn,10pt]{wlscirep}
\title{A comparative study between two models of active cluster crystals}

\author[1,*,+]{Lorenzo Caprini}

\author[2,+]{Emilio Hern\'andez-Garc\'\i a}
\author[2,+]{Crist\'obal L\'opez}
\author[3,+]{Umberto Marini Bettolo Marconi}
\affil[1]{Gran Sasso Science Institute (GSSI), Via. F. Crispi 7, 67100 L'Aquila, Italy.}
\affil[2]{IFISC (CSIC-UIB), Instituto de F\'{\i}sica Interdisciplinar y Sistemas Complejos,
Campus Universitat de les Illes Balears, E-07122 Palma de Mallorca, Spain.}
\affil[3]{Scuola di Scienze e Tecnologie, Universit\`{a} di Camerino - via Madonna delle Carceri,
62032, Camerino, Italy.}

\affil[*]{lorenzo.caprini@gssi.it}

\affil[+]{these authors contributed equally to this work}

\usepackage{graphicx} \usepackage{bm,ulem,color,textcomp}
\usepackage{amsmath,amsfonts,amssymb} 

\usepackage{subfig}
\usepackage{graphicx}

\usepackage[justification=justified, format=plain]{caption}

\usepackage{comment}

\usepackage{soul} 

\begin{abstract}
We study a system of active particles with soft repulsive
interactions that lead to an active cluster-crystal
phase in two dimensions. We use two different modelizations of the active force -
Active Brownian particles (ABP)  and Ornstein-Uhlenbeck
particles (AOUP)  - and focus on analogies and differences
between them. We study the different phases appearing in the system,
in particular, the formation of ordered patterns drifting in space
without being altered.
 We develop an effective description which
captures some properties of the stable clusters for both ABP
and AOUP.
As an additional point, we confine such a system in a large channel,
in order to study the interplay between the cluster crystal phase and
the well-known accumulation near the walls, a phenomenology typical of active particles.
For small activities, we find clusters attached to the walls and deformed, while for large
values of the active force they collapse in stripes parallel to
the walls.

\end{abstract}

\begin{document}


\flushbottom
\maketitle
\thispagestyle{empty}

\section{Introduction}

Active particle models have been introduced to investigate the
dynamics of individuals or objects able to convert energy
from the environment into directed motion~\cite{ramaswamy2010mechanics, marchetti2013hydrodynamics, bechinger2016active, fodor2018statistical}.
The active systems  most studied so far are
bacteria \cite{berg2008coli}, protozoa \cite{blake1974mechanics}, spermatozoa \cite{woolley2003motility}, cells, living tissues \cite{poujade2007collective}, actin filaments \cite{kohler2011structure}, active nematics \cite{guillamat2016control},
and the so-called motor-proteins \cite{sanchez2012spontaneous}.
Recently, active microswimmers have been synthesized in
labs \cite{bechinger2016active}. Typical examples are the Janus
particles \cite{takatori2016acoustic},
spherical objects containing two faces of distinct properties, such as
hydrophobicity and hydrophilicity.
Such an asymmetry produces a self-propulsion in a given direction.
The ``activation'' of a complex microswimmer could have
important applications, for instance in drug delivery:
a propulsion mechanism could enhance and direct the transport
process with a consequent increase of its efficiency.
Alternatively, some colloids could be activated by light or magnetic fields \cite{ebbens2010pursuit}.

A large interest in the polymers community resides in the synthesis and
study of star polymers \cite{LikosPRL1998} or
dendrimers \cite{madaan2014dendrimers, Mladek2008}, which have highly
branched structures and thus endowed with several interesting
characteristics such as globular, void-containing, shapes which
make them suitable for the delivery of anticancer drugs and
imaging agents.
As a consequence of the presence of cavities and channels in
their interior, the interaction between different dendrimers or
star polymers in solution could be modeled by means of
soft-core interactions, which do not prevent the overlap
between particles. In particular, it may be described by
Generalized Exponential Model (GEM) potentials whose properties
have been reviewed by Likos~\cite{Likos2001a,
BookLikosSciortinoZaccarelli}. Following this, we
adopt a coarse-grained model which replaces a suspension of
complex polymers with overlapping spherical particles. 
Remarkably, under equilibrium conditions, these particles may
bind together to form clusters and these, in turn, may organize
periodically to form cluster crystals \cite{caprini2018cluster,
Delfau2015, CoslovichIkeda2013, Likos2007, Mladek2006}.
Because of the peculiarities of two-dimensional systems
\cite{strandburg1988twodimensional}, this would not be a true
crystal phase with long-range positional order, but it displays
a clear inhomogenous and periodic distribution of particles,
much more clustered and ordered than the homogenous state found
at high temperature. Such an interesting aggregation phase
occurs also in the presence of other soft-core potentials
besides the GEM ones, such as the ultra-soft core potentials
used for low-temperature bosons \cite{Cinti2014,
diaz2015monodisperse, wang2019melting} and vortices in
superconductors \cite{varney2013hierarchical}.

The self-propulsion is often modeled by means of
an effective stochastic  force. Perhaps, the simplest model
introduced in the literature to describe the behavior of some
bacteria populations is represented by the discrete
run-and-tumble particle dynamics \cite{tailleur2008statistical,
nash2010run}. Recently, suitable descriptions in terms of
continuous stochastic processes have received large attention
for the possibility of applying well-known tools of statistical
mechanics towards the development of the thermodynamic of
active microswimmers. In this framework, we mention the active
Brownian particles (ABPs)~\cite{ten2011brownian,
romanczuk2012active} and the active Ornstein-Uhlenbeck
particles (AOUPs)~\cite{szamel2014self, marconi2015towards,
marconi2016velocity, szamel2015glassy} models. Regarding the
observed phenomenology, there is a strong resemblance between
ABP and AOUP, but they also display important differences: a)
the ABP  active force has a fixed strength whereas its
orientation fluctuates in a diffusive fashion, while in the
AOUP there is no such a constraint and the propulsive force
fluctuates both in strength and direction; b) the  correlations
are of Gaussian type  in the AOUP and non-Gaussian in ABP, in
spite of the fact that they share the same two-time
self-correlation of the active force~\cite{caprini2019activity,
das2018confined}. 
 So far, the practical
consequences of these differences have not been completely elucidated. One can say that in the majority of cases the two models
display a similar phenomenology, whose characteristics are:
\begin{enumerate}

\item[i)]  In the absence of external forces some single particle properties
such as the diffusion and the mean square displacement has the same form in ABP \cite{ten2011brownian, sevilla2015smoluchowski} and AOUP \cite{Gardiner}.

\item[ii)]  Both models undergo the so-called motility induced phase separation \cite{fily2012athermal, cates2015motility, gonnella2015motility, ginot2018aggregation}. While such a topic has been well studied in the ABP case, with steric interactions \cite{digregorio2018full, bialke2015active, speck2016collective, mandal2019motility}, in the presence of an attractive component of the potential \cite{redner2013reentrant, buttinoni2013dynamical} and for more complex interactions \cite{pu2017reentrant, fischer2019aggregation}, the result with AOUP model appeared only in \cite{fodor2016far}.

\item[iii)]  In the two models particles accumulate in the proximity of the walls
\cite{maggi2015multidimensional, wensink2008aggregation}. In particular,
the wall induces a non-uniform density profile decaying with a characteristic length-scale
\cite{caprini2019active, yan2015force, yan2018curved}.

\item[iv)] The dynamics in the presence of a convex, radial
    and non-harmonic potential shows particle
    accumulation far from the minimum of the potential
    \cite{caprini2019activity, takatori2016acoustic,
    dauchot2019dynamics} when the activity is large, in
    such a way that the particle distribution is not
    Boltzmann-like \cite{marconi2017heat,
    malakar2019exact}.

\end{enumerate}
The extension of the ABP and AOUP models of activity to
particles interacting with soft potentials is considered in
the present paper. We analyze how the morphological properties
of the system of particles depend on the specific modeling of
the active force and if some features of the
dynamics are model-independent.
What happens if the non-equilibrium forcing
is replaced by a colored
noise term as in the AOUP model?
As an example, it has been found with ABP driving that for some values of the activity parameter the clusters
deform into rings with an empty interior \cite{Delfau2017}, at variance with
the equilibrium situation where clusters are compact.
We will see here that such an empty-cluster crystal phase does not appear in the AOUP description.

Hereafter, we shall study in detail some properties, such as phase
diagram and cluster size, stressing similarities and
differences between the AOUP and ABP models. Moreover,
motivated by microfluidics applications, we confine the system in
large channel to evaluate the long-range influence of the
walls on the active cluster-crystal-phase.

After introducing the active ultrasoft
model in the ABP and AOUP versions,
we present a numerical
study, for high enough density, displaying
a traveling cluster-crystal aggregation phase. The role of the
active force is elucidated, determining at first the size of
the cluster and then the occurrence of unstable regions, where
clusters shrink and reform.
Then, an effective description of the
the system is developed, with the aim of describing both the
microscopic dynamics of a particle within a cluster and the
global dynamics of the pattern.
Finally, we confine the system in a
large channel to explore the interplay between the
cluster-crystal phase and the accumulation near the walls.
In the last Section we summarize the results discussing future
perspectives in the conclusive section.

\section{Model}\label{sec:model}
Polymers are often described as large complex structures
having many internal degrees of freedom, but in some cases
it is not necessary nor possible to take into account their internal properties.
For instance, when the resolving power of instruments is low some details of their structure
can be disregarded and the polymers can be assimilated to diffusing objects.
It is coherent with such an approach to represent the effective interaction among different
polymers by a pair-wise potential which depends on the coordinates of their centers of mass.
  In some cases, such as for dendrimers,
this interaction is repulsive and of the soft-core type \cite{Likos2001a}, basically due to non-diverging potentials
which do not prevent the overlap among such complex
structures.
An appropriate model for the effective interaction
is the so-called GEM-$\alpha$ potential which reads:
\begin{equation}
\label{eq:softcorepotential}
V(\{ \mathbf{x)} \} ) \propto \sum_{i<j} \phi(\bm{x}_{ij}), \qquad \phi(\bm{x}_{ij})
=\epsilon ~ e^{-(|\bm{x}_{ij}|/R)^{\alpha}} \,,
\end{equation}
being $\alpha$ a positive real number, and $\{|\bm{x}_{ij}|\}$ are
the relative distances between the pairs of particles. We study
a system of point particles which self-propel and interact with
this potential in two dimensions. Depending on the stiffness of
the potential (in particular, $\alpha>2$ is needed), the
passive system (i.e. without the active self-propulsion) shows
the occurrence of a peculiar aggregation phase at equilibrium
\cite{Likos2007, Mladek2006, Mladek2008, Delfau2015}: particles form stable clusters,
which arrange into a periodic configuration. In two dimensions
the triangular or hexagonal lattice is the only stable pattern,
occurring for small enough temperature $T$
at a large density.
The number of
particles, $N_c$, of each cluster depends on the global density
and on the typical interaction length, $R$, while the typical
inter-cluster distance is determined just by $R$. When the
particles are no longer point-like but of small finite size
\cite{caprini2018cluster, Glaser2007, Ziherl2011, Shin2009} similar properties are found, with a
new low-temperature phase -the so-called crystal
cluster-crystal phase- in which particles inside the clusters
also show an ordered structure. The cluster-formation
phenomenon despite the repulsive interaction between all the
particles can be physically interpreted by considering the
force balance between the intra-cluster repulsion (the force
felt by particles inside clusters) and the inter-cluster
effective interaction, that is, the force exerted by
neighboring clusters \cite{Delfau2015, caprini2018cluster}.
The increasing of temperature enlarges the typical size of
clusters, destroying any structure for $T$ sufficiently large.

As mentioned in the introduction, these complex structures
could be ``activated'' by chemical reactions or biological
mechanisms taking place inside or on the surface of such a
complex microswimmer, so that each individual self-propels in a
preferential direction in the same way as a simple rod-like
active particle in the absence of any structure.
The occurrence of a driving velocity could deform the structure
of the polymer, destroying its circular symmetry and
altering the shape of the effective interaction given by
Eq.\eqref{eq:softcorepotential}. But when the speed induced by
the self-propulsion is smaller than the typical velocities of
the ``microscopic'' components of the polymer (for instance,
the arms of star polymers) then the structural changes of the
polymer shape would be negligible and our description in
terms of point particles with effective interaction and
self-propulsion remains valid. We restrict in the
following to this regime.

Self-propulsion can be modeled by means of a force
vector, applied on the center of mass of each microswimmer with
dynamics
 independent on the particle position \cite{bechinger2016active, ramaswamy2010mechanics}.
In particular, we consider a two-dimensional system of $N$
interacting active particles, whose dynamics is described by
over-damped Langevin equations for the positions,
$\mathbf{x}_i$, of each particle:
\begin{equation}
\label{eq:xdynamics}
\gamma\dot{\mathbf{x}}_i= \bm{F}_i  + \sqrt{2 \gamma T}\,\bm{\eta}_i
+ \gamma\mathbf{f}_i \ ,
\end{equation}
where $\bm{F}_i=-\nabla_i V$ is the total force exerted on the
particle $i$ by the rest of the particles due to the repulsive
potential $V$ (as given by Eq.\eqref{eq:softcorepotential}). The
term $\sqrt{2\gamma T}\boldsymbol{\eta}_i$ represents the
effect of a thermal bath at temperature $T$ where active
particles are immersed, $\bm{\eta}=(\eta_x, \eta_y)$ is a
two-dimensional Gaussian white noise vector with zero average
and correlations $\langle \bm{\eta}_i
(t)\bm{\eta}_j(t')\rangle= \mathbb{I} \delta_{ij}\delta(t-t')$,
with $\mathbb{I}$ the 2d identity matrix. The constant $\gamma$
is the drag coefficient. The last term, $\mathbf{f}_i$,
models the self-propulsion mechanism of the microswimmers, and
here is where the ABP and AOUP approaches enter into the
modelling. In the ABP, $\mathbf{f}_i$ is given by a vector of
fixed norm: $\mathbf{f}_i= U_0 \mathbf{\hat{n}}_i$, where
$\mathbf{\hat{n}}_i=(\cos{\theta_i}, \sin{\theta_i})$ is a unit
vector whose angle, $\theta_i$, evolves as a Wiener process
\begin{equation}
\label{eq:ABPactiveforce}
\dot{\theta}_i = \sqrt{2D_r}\xi_i \,.
\end{equation}
The constant $D_r$ is the rotational diffusion coefficient,
while $\xi_i$ is a Gaussian white noise with zero average and
correlations $\langle \xi_i (t)\xi_j(t')\rangle=
\delta_{ij}\delta(t-t')$. Instead, in the AOUP model
$\mathbf{f}_i$ is a noise vector whose components evolve as
independent Ornstein-Uhlenbeck processes:
\begin{equation}
\label{eq:AOUPactiveforce}
\tau\dot{\mathbf{f}}_i = -\mathbf{f}_i +\sqrt{2D_a}\mathbf{w}_i \,,
\end{equation}
where $\tau$ is a correlation time, $D_a$ an effective
diffusion constant characterizing the active force, and
$\mathbf{w}_i$ is a Gaussian white noise vector with zero
averages and correlations $\langle \mathbf{w}_i
(t)\mathbf{w}_j(t')\rangle= \mathbb{I}
\delta_{ij}\delta(t-t')$.
We point out the relevance of the ratio, $D_a/\tau$, i.e. the variance of $\mathbf{f}_i$,
whose square root gives the typical average value of the active force norm.

Despite the differences between ABP and AOUP models, a
connection line between them has been already explored in
\cite{das2018confined}. We remark that AOUP has been
originally introduced as a simplification of ABP in order to
capture its phenomenology with the aim of making analytical
predictions \cite{fily2012athermal}.
Subsequently, AOUP was also used to describe the complex behavior of a passive object
immersed in a bacterial bath \cite{maggi2014generalized, maggi2017memory, steffenoni2016interacting}.
Anyway, AOUP is the simplest Gaussian model which displays
at long times the same average and two-time activity-activity correlation
function as the ABP.
In particular, this correlation in the ABP
case has an exponential form, which in two dimensions reads \cite{Farage2015}:
\begin{equation}
\label{eq:activityactivityABP}
 \langle  \mathbf{f}_i(t) \mathbf{f}_j(0) \rangle =
 U_0^2 \langle \hat{\mathbf{n}}_i(t) \hat{\mathbf{n}}_j(0) \rangle =
 \delta_{ij}\mathbb{I} \frac{U_0^2}{2} e^{-D_r t} \,.
\end{equation}
 Instead, in the AOUP model, such a correlation is \cite{Gardiner}
\begin{equation}
\label{eq:activityactivityAOUP}
 \langle  \mathbf{f}_i(t) \mathbf{f}_j(0) \rangle =
 \delta_{ij}\mathbb{I} \frac{D_a}{\tau} e^{-t/\tau} \,.
\end{equation}
Setting $\tau=1/D_r$ and $2 D_a/\tau= U_0^2$
the two-time activity-activity correlation
of the AOUP dynamics coincides with Eq.\eqref{eq:activityactivityABP} \cite{caprini2019activity},
and the average is $ \langle  \mathbf{f}_i(t) \rangle =0$ in both cases.

As commented in the introduction two main differences appear
between ABP and AOUP active forces: the fluctuating norm of
the AOUP  force \cite{das2018confined}, and the higher order
correlations \cite{sevilla2014theory} which make the ABP force
non-Gaussian. We will explore in the following if these
differences (and which) are relevant for active cluster
crystals.

\section{Numerical Results}\label{sec:numericalresults}

In this Section, we numerically explore the dynamics of
Eq.~\eqref{eq:xdynamics} for a suspension of $N$ interacting
active particles in a two-dimensional box of size $L$ with
periodic boundary conditions. We implement the Euler-Maruyama
algorithm \cite{toral2014stochastic} using both ABP and AOUP active forces, given by
Eq.\eqref{eq:ABPactiveforce} and \eqref{eq:AOUPactiveforce},
respectively. We choose a soft-core potential of the
GEM-$\alpha$ type with $\alpha=3$, which displays the
cluster-crystal aggregation phase in the passive case ($U_0=0$)
at sufficiently low temperature and large enough particle density \cite{Delfau2015}. Throughout this paper
we take $R/L=10^{-1}$, $\gamma=1$ and $\epsilon=1$. Since for
several active systems of interest  the effective diffusion
due to the active forces is much larger than the one
due to the thermal diffusion, hereafter we fix $T=(U_0^2 D_r/\gamma
)\, 10^{-4}$.

\begin{figure}[!t]
\centering
\includegraphics[clip=true,width=0.7\columnwidth,keepaspectratio]{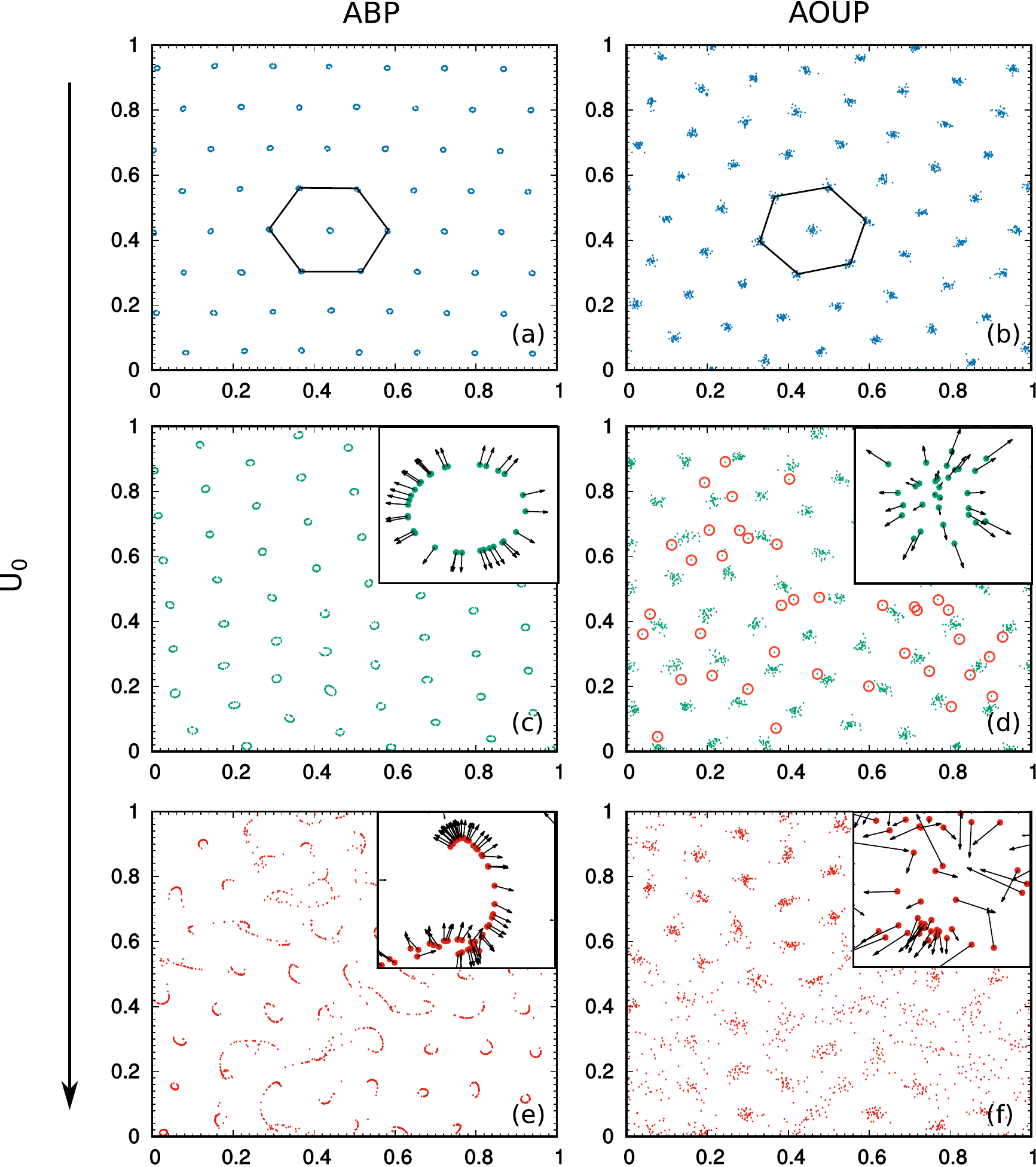}
\caption{Snapshot of the configurations in the plane $xy$ of a system of
$N=2\times 10^3$ particles interacting with the GEM-3 potential for both
ABP (left column) and AOUP (right column) active forces.
Panels (a) and (b) are obtained with $U_0=2.75$, panels (c) and (d) with
$U_0=3.75$ and panels (e) and (f) with $U_0=4.5$.  Other parameters: $\gamma=1$,
$D_r=1$, $L=1$, $R=10^{-1}$, $\epsilon=1$, $T=10^{-4} U_0^2/D_r \gamma$. In the AOUP case the
parameters used are $D_r=1/\tau$ and $U_0^2=2 D_a/\tau$. Black lines in
panels (a) and (b) are eye-guides to show the hexagonal pattern.
Graph (c), (d), (e) and (f) display an inset around a small square region
around a cluster. In the insets, the black arrows have a length proportional to the
active force on each particle. In panel (d), isolated particles not belonging to clusters
are highlighted by red circles.
}
\label{fig:Snapshot}
\end{figure}

\begin{figure}[!t]
\centering
\includegraphics[clip=true,width=0.9\columnwidth,keepaspectratio]{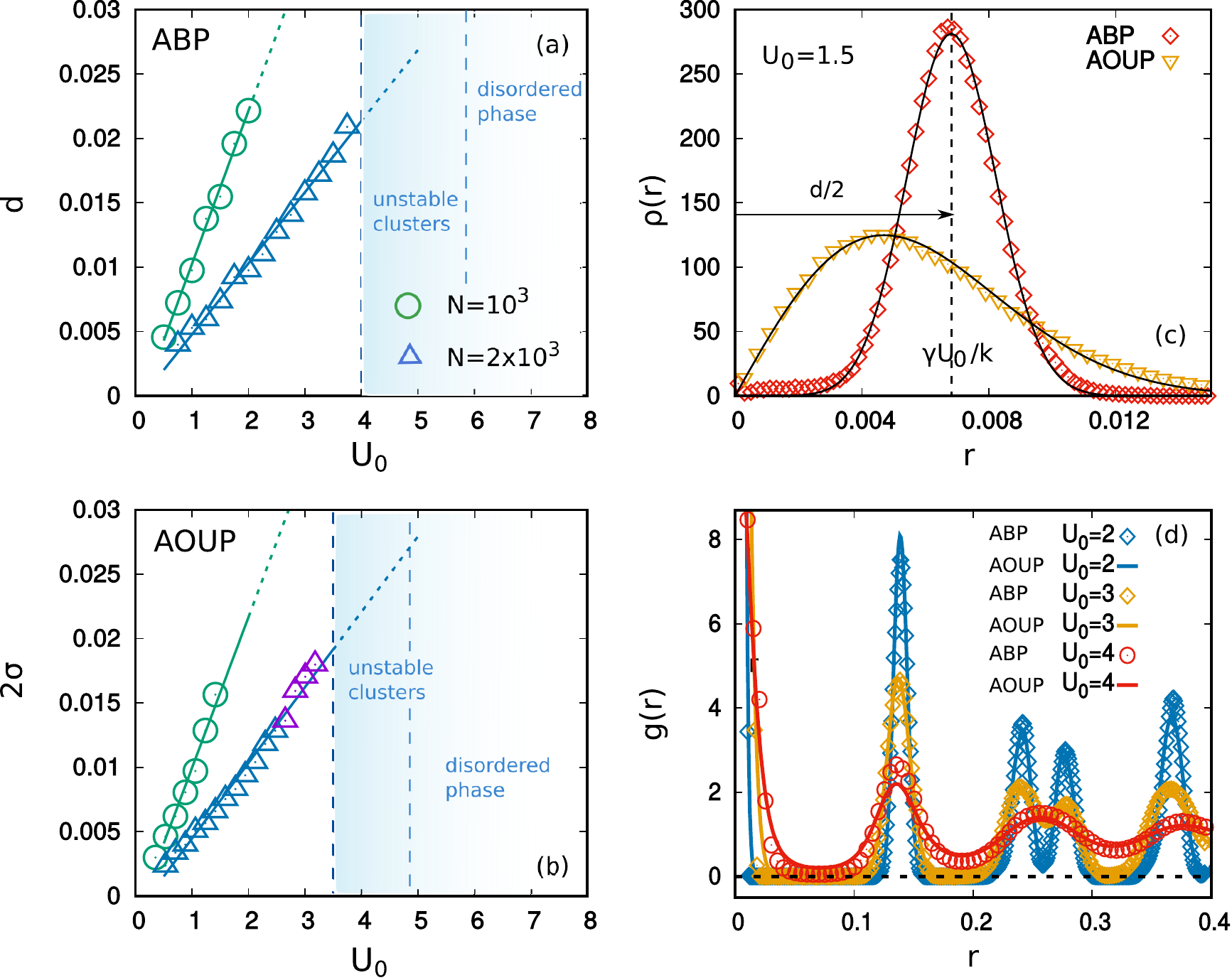}
\caption{Cluster size as a function of $U_0$ for two different values of the density:
green  and blue  for $N=10^3, 2\times 10^3$, respectively, at fixed $L$.
Other parameters as in Fig. \ref{fig:Snapshot}.
Continuous lines are obtained by linear numerical fits. In the ABP case (Panel (a))
we plot the diameter, $d$, of the ring cluster, while the AOUP case (Panel (b)) we
show $2\sigma$. In the
AOUP case the label $U_0$ is a short-cut for $\sqrt{2 D_a/\tau}$. The appearance of
configurations which exchange particles among clusters is indicated by violet symbols.
Blue shades denote the occurrence of unstable clusters in the case $N=2\times10^3$,
while the same analysis for $N=10^3$ is not shown for the sake of clarity.
The left vertical dashed line indicates the
smallest value of the active force which displays the
instability of at least one cluster in the steady state.
The right vertical dashed line shows the beginning of the disordered phase.
In panel (c) we show the radial probability density,  $\rho(r)$.
Red diamonds refer to ABP, yellow triangles to  AOUP.
Black lines are the result of a numerical fit.
In panel (d) we plot the pair correlation function, $g(r)$,
for different values of the active force both for ABP and AOUP dynamics, as shown in the legend.
}
\label{fig:clustersize}
\end{figure}

In Fig.\ref{fig:Snapshot} we show long-time patterns ( $t/\tau\approx 10^2$) for different values of $U_0$
for ABP (left column) and AOUP (right column).
At such a low temperature,
both ABP and AOUP form clusters
arranged into a hexagonal pattern (see panels (a), (b), (c) and
(d)). This scenario resembles the one of passive particles
($U_0=0$) \cite{Delfau2015, caprini2018cluster}.

Notwithstanding that for low values of $U_0$ the active force
does not change the static macroscopic properties of the
pattern, i.e. the hexagonal cluster arrangement, strong
differences appear at the dynamical level. While clusters
occupy stable equilibrium positions in the passive case
($U_0=0$), without displaying any macroscopic motion, this
situation changes in the active case ($U_0 >0$). The active
forces produce a macroscopic coherent motion of the pattern,
which maintains the hexagonal cluster-crystal phase. Clusters,
stuck in the hexagonal pattern, drift persistently before
changing direction after a time which grows as $1/D_r$. In
movie 1 of the Supplementary Information, we compare the time
evolution of systems with $N=2\times 10^3$ particles for ABP
and AOUP. Despite the total active force acting on each cluster
is directed randomly in space (green arrows in movie 1),
clusters move coherently in one direction maintaining the
hexagonal arrangement. This global drift follows the average
active force of the whole system (black arrow in movie 1). The
decreasing of $D_r$ enlarges the persistence of the pattern
dynamics, for both cases, without producing any significant
change on the particles configurations. 
Traveling
crystals \cite{menzel2013traveling} occur also for highly
packed suspensions of self-propelled particles interacting
by hard-core potentials, i.e. interactions diverging at $r=0$
in such a way that a finite size is attributed to the
particles. Experimental evidence of this effect has been
recently studied by means of a suspension of micro-disks
subjected to vertical vibrations
\cite{briand2018spontaneously}.
The use of density functional theory predicts such a
phenomenology and, in particular, the transition to rhombic,
quadratic, and lamellar patterns as the active force is
increased \cite{menzel2013traveling,menzel2014active}. Such
transitions do not occur in our system, where the only stable
pattern is the hexagonal one. This statement is confirmed
by movie 1 in Supplementary Information and by the study
of the pair correlation, $g(r) = \left\langle \sum_{i \neq 0}
\delta\left(x - x_i \right) \right\rangle  L^2/N$, where a target particle is at the origin, the sum runs over the other
particles, and
the brackets indicate  
a circular average over positions $\mathbf{x}$ with the same
modulus $|\mathbf{x}| = r$. In Fig.\ref{fig:clustersize}d), we
compare the $g(r)$ for several values of the active force both
for ABP and AOUP dynamics, revealing the occurrence of the
typical peaks of a hexagonal pattern, indicating the
presence of first, second, third, etc. neighbors at distances
$1$, $\sqrt{3}$, $2$, $\sqrt{7}$, ... times the basic
periodicity of the pattern. This periodicity does not change
with $U_0$ and remains at the value determined by this
potential in the passive case
\cite{Delfau2015,caprini2018cluster}, namely $1.4~R\approx
0.14$. The increasing of $U_0$ leads to larger and lower peaks
towards the occurrence of the liquid shape of the $g(r)$. The
presence of the initial peak near $r\sim 0$ is due to the $N_c$
particles belonging to each cluster. We remark that the $g(r)$
remains essentially unchanged when computed at different
times during the motion (this is not shown in the figure). 

The specific modeling of the active force influences the structure of a
single cluster as shown in the insets of
Fig.~\ref{fig:Snapshot}, panels (c), (d), for ABP and AOUP, respectively.
In particular, ABP's arrange in ring-like clusters
while AOUP's form Gaussian-like clusters as in the
passive case ($U_0=0$).
In Fig.\ref{fig:clustersize}c) the radial probability distribution, $\rho(r)$, of a
single cluster is numerically measured.
 $\rho(r)dr$ is the fraction
of particles of a cluster which are at a distance between $r$
and $r+dr$ from its center.
In particular, $\rho(r)=2\pi r p(\mathbf{x})$, where $p(\mathbf{x})$ is the
single-particle probability density for a specific particle to
be at $\mathbf{x}$, which turns out to depend only on the
norm $r=|\mathbf{x}|$.

In the AOUP case,
$p(\mathbf{x})$ is fitted by a Gaussian centered at
$\mathbf{x}=0$, while in ABP case $p(\mathbf{x})$ is close to a
Gaussian in $r=|\mathbf{x}|$ centered at a typical radius
larger than zero, and displays a smaller variance. For both   ABP
and AOUP  the angular location of the particles
in the clusters, i.e. the angular coordinates $\phi_i$ computed
for each particle $i$ with respect to the center of its cluster, is
approximatively equal to the orientational angle of the active
force, $\theta_i$. This feature is illustrated in the insets
 of Fig.\ref{fig:Snapshot}c) and d), where black
arrows represent the vectors of the active forces for each
particle, which point radially with respect to the center of
the cluster. Despite the apparent analogies  between the clusters obtained with the AOUP model
and the passive systems \cite{Delfau2015} (both are
non-empty), we
note the difference between these two cases: in
the large persistence regime, $D_r \ll \gamma$
an AOUP particle is stuck to its radial position
with respect to the center of its cluster only for a time $\sim
\tau$. After this correlation time, the active force is
consistently modified and, as a consequence, a large variation
in the radial coordinate of the particle occurs. In fact,
the non-existence of ring-like clusters in the AOUP
is due to the fluctuating norm of the active force at variance with the ABP where it remains constant. This fact is confirmed by simulations (not shown) where
a force with the angular time-dependence of the AOUP but
a fixed norm is used in analogy with \cite{das2018confined}.
 In this case, empty clusters are obtained.
Also, for some particular initial conditions in Eq.
(\ref{eq:AOUPactiveforce}), ring-like clusters are observed for
the standard AOUP, but only as a short-lived (for a time of the
order of $D_r^{-1}$) transient state.

In Fig. \ref{fig:clustersize} we study the size of a single
cluster, i.e. the average diameter of the ring in the ABP case
(Panel (a)) or twice its standard deviation in the AOUP case
(Panel (b)), as a function of $U_0$ ($=\sqrt{2D_a/\tau}$ in the
AOUP case). In particular, linear scaling of the standard
deviation with $U_0$ emerges for both types of active drivings
and its slope depends on the mean number of particles in each
cluster $N_c$: larger values of  $N_c$ produce larger
confinements. Such a  feature is due to the stronger repulsion
from the neighboring clusters, which is more important than the
intracluster repulsion, and needs the contribution from the
active forces to get balanced at a particular ring
radius~\cite{Delfau2017}. We remark that such size does not
depend on the persistence time of the active force, $\tau$. A
second difference between the two models emerges at large
values of $U_0$ (or of $\sqrt{2 D_a/\tau}$). As shown by the
comparison between Panels (c) and (d) of
Fig.\ref{fig:Snapshot}, at some values of $U_0$  the AOUP
reveals a different phase which does not occur in the ABP
model: In panel (d) we observe that some particles, which are
marked by red circles, leave their clusters, eventually joining
other ones, even if the cluster-crystal phase is preserved.
Quantitatively, this particle exchange is a rare event and does
not affect the cluster-size as shown in
Fig.\ref{fig:clustersize}. The number of particles which
migrate from a cluster to another, without destroying the
stable hexagonal pattern, grows with increasing $\sqrt{2
D_a/\tau}$ in the AOUP but is always a very small fraction of
$N_c$. For the ABP case, the exchange of particles could
eventually occur only for larger values of $T$ as found in
\cite{Delfau2017}, but is always independent of $U_0$.

A further increase of $U_0$ leads for both models to the
instability of some clusters. In such cases, clusters disappear
and immediately reform. The number of unstable clusters
increases as $U_0$  grows, without destroying the global order
of the pattern. While in the AOUP case clusters simply blow
up, in the ABP case we can clearly see the deformation of the
ring-like aggregates into lines after which the cluster can
reform. Then, the growth of $U_0$  produces the formation of
continuous flows of particles while other groups of particles
arrange in the hexagonal pattern (Fig.\ref{fig:Snapshot}c) and f)).
 In other words, we observe the coexistence of the
cluster-crystal phase with the disordered phase. Finally, the
formation of any stable structure for both models is prevented
by a further increase of $U_0$, a regime which is not under
investigation in the present manuscript.

The comparison between ABP and AOUP shows the
weaker stability of the cluster-crystal phase in the latter,
due to the particle-exchange
mechanism which does not occur in the ABP case. In addition,
clusters start to break down for AOUP at smaller values of $U_0$
as shown in Fig.\ref{fig:clustersize} (for $N=2\times 10^3$).
 The
disordered phase in  ABP  appears for larger values of
$U_0$ than for  AOUP. We remark that in both cases the
active force plays a role which resembles the one of an
effective temperature since its increasing induces a transition
towards the disordered phase. In any case, the role of the
active force reveals other interesting features, like the
the coherent motion of the pattern and the shape of a single
cluster, which cannot be understood in terms of an effective
temperature approach.

\section{Effective description}\label{sec:effectivedescription}

In this Section, we elaborate an effective description of the
dynamics of the system in the cluster-crystal phase by
separating the dynamics of the single particle from the
dynamics of the clusters. This has already been done for
equilibrium hard-core passive repulsive particles in
\cite{caprini2018cluster} and is based on the fact that the
typical distance between clusters remains approximately
constant. As commented before, such assumption holds also
in the presence of the active force (see the movie in
Supplementary Information). These observations allow us to
separate the effective dynamics of the single particle from the
effective dynamics of the clusters. In particular, the
effective equations for the $i$-th particle in the $j$-th
cluster reads
\begin{equation}
\label{eq:effectivedynamicsposition}
\gamma\dot{\mathbf{x}}_i^{(j)} = \boldsymbol{F}_{eff}\left( \mathbf{x}_i^{(j)} - \mathbf{R}^{(j)} \right)  +\gamma \mathbf{f}_i^{(j)} + \sqrt{2\gamma T} \boldsymbol{\eta}_i^{(j)} \,,
\end{equation}
where $\mathbf{F}_{eff}$ is the effective confining force due to
the neighboring clusters \cite{Delfau2017}.
The mean cluster
positions, $\mathbf{R}^{(j)}$, are located on a hexagonal structure
as in the purely Brownian case ($U_0=0$) since the
inter-cluster distance does not change significantly by the
presence of the active force.
The complex dynamics of interacting microswimmers is approximated by a set
 of independent particles in the presence of a grid of confining potential wells.
The validity of this approximation has been discussed in \cite{Mladek2006, Delfau2015,caprini2018cluster}
in the passive case, and basically follows by a Taylor expansion of the GEM-$\alpha$
 potential truncated at the second order, since the inter-cluster distance
is always larger than the typical cluster-size in the cluster
crystal phase. Within this approximation, particles belonging
to the same cluster are treated as independent and only
experience the effective force generated by the particles in
the neighboring clusters \cite{Delfau2017}, which is described
by the linear shape, $\mathbf{F}_{eff}\left(\mathbf{x}\right)
\approx -k \left( \mathbf{x}-\mathbf{R}^{(j)}  \right)$.
We stress that the same equations are obtained if using
interaction potentials different from GEM, the only difference
being the particular value of the spring constant $k$. At
variance with the equilibrium case where the pattern does not
move and each cluster fluctuates around its equilibrium
position, in the presence of the self-propulsion the positions
$\mathbf{R}^{(j)}$ change. However, since particle relative
distances remain constant within the cluster we shall neglect
this  movement and only consider the particle dynamics inside
each cluster and study a system of independent particles
confined in a harmonic well in the presence of the
self-propulsion.
 Taking the center of the cluster ($\mathbf{R}^{(j)}=\mathbf{0}$) as the origin of coordinates, the effective particle dynamics reads
\begin{equation}
\label{eq:xeffectivedynamics}
\gamma\dot{\mathbf{x}}= -k\bm{x}  + \sqrt{2  \gamma T}\,\bm{\eta} + \gamma\mathbf{f} \,,
\end{equation}
where we dropped the indices $i$ and
$j$ for the sake of simplicity.

In the AOUP case Eq.~\eqref{eq:xeffectivedynamics} is
linear, so that we can solve its associated  Fokker-Planck equation
for the steady state joint probability distribution function, $f(\mathbf{x},\mathbf{f})$, which reads \cite{caprini2019active}:
\begin{equation}
\label{eq:fxfa_AOUP}
f(\mathbf{x}, \mathbf{f})= \mathcal{C}
\exp{\left(- \dfrac{k}{2}\dfrac{\left|\mathbf{x}\right|^2}{D_a\gamma} \dfrac{D_a \Gamma}{D_a+\Gamma T/\gamma}  \right)}
\exp{\left(-\dfrac{\tau \Gamma}{2D_a}\left| \mathbf{f} - \dfrac{k}{\gamma} \dfrac{\Gamma D_a}{D_a+T/\gamma} \mathbf{x}  \right|^2  \right)} \,,
\end{equation}
where $\mathcal{C}$ is a normalization and $\Gamma=1+k \tau/\gamma$ is a numerical factor
 which depends on $k\tau/\gamma \gg1$, i.e. the ratio between the correlation
 time of the activity, $\tau$, and the relaxation time, $\gamma/k$, due to the harmonic potential.
The result is a multivariate Gaussian distribution with
non-zero correlations between each component of $\mathbf{x}$
and $\mathbf{f}$. In other words, a non-zero conditioned
first moment of the spatial distribution appears, so
 that $\langle \mathbf{x} \rangle \propto \mathbf{f}$, meaning that particles
prefer to spend their life far from the minimum of the potential in a fixed position determined by the value of $\mathbf{f}$.
Since the active force is an Ornstein-Uhlenbeck process, $\mathbf{f}$ can explore
 large values depending on its variance, $D_a/\tau$, even if the most probable values remains $\mathbf{f}=0$.
Instead, its persistence time, $\tau$, rules how long the particle remains close to
 the particular position determined by the value of $\mathbf{f}$.

Integrating out the active force in Eq.\eqref{eq:fxfa_AOUP}, we can easily find the probability
density for the position of a given particle, $p(\mathbf{x})$, which reads:
\begin{equation}
\label{eq:AOUPdensity}
p(\mathbf{x}) =  \mathcal{N} e^{-\dfrac{k (x^2+y^2)}{2 \gamma} \dfrac{\Gamma}{ D_a+\Gamma T /\gamma} } \,,
\end{equation}
where $\mathcal{N}$ is a normalization factor.
Formula \eqref{eq:AOUPdensity} implies that the clusters have a Gaussian
shape with  $\langle x\rangle=\langle y\rangle =0$ and defines an effective temperature of the system \cite{szamel2014self}:
\begin{equation}
\label{eq:varianceAOUPharmonic}
T_{e} =k \langle x^2 \rangle = k\langle y^2 \rangle=  D_a \gamma  \left(\dfrac{T}{\gamma D_a} + \dfrac{1}{1+k\tau/\gamma}   \right) \approx  \dfrac{D_a \gamma}{1+k\tau/\gamma} .
\end{equation}
 The last approximation holds
if $T \ll \gamma D_a$, the regime considered in this manuscript.
Note that the result for $T_e$  is in agreement with the scaling with $\tau$
recently observed in a dense suspension of active particles with hard-core
repulsive interaction, specifically Lennard Jones potentials \cite{mandal2016active, mandal2019extreme}.
We also note that Eq.\eqref{eq:varianceAOUPharmonic} approaches
 $T+\gamma D_a$, in the equilibrium limit, $\tau \to 0$, i.e. the effective temperature due
to the joint effect of self-propulsion and thermal noise for a
free microswimmer \cite{ten2011brownian, winkler2015virial, kurzthaler2016intermediate}.
The Gaussianity of the density
agrees with the shape of the clusters obtained in
Fig.~\ref{fig:clustersize}c). Moreover, the standard
deviation of the above distribution, obtained as  the square
root of Eq.~\eqref{eq:varianceAOUPharmonic}, confirms the
linear cluster-size scaling with $\sqrt{2 D_a/\tau} = U_0$,
numerically measured in Fig.~\ref{fig:clustersize} (b). We note
that the authors of ~\cite{caprini2019activity}
showed that significant
anharmonicity of the trap (for instance
$\mathbf{F}_{eff}=-\nabla U(\mathbf{x})$, with $U(\mathbf{x})\propto
|\mathbf{x}|^{2n}$ and $n\geq 2$) would lead, under AOUP dynamics, to a
``delocalization'' phenomenon, in which particles accumulate
far from the minimum of the potential,
 displaying a non-Boltzmann distribution
\cite{marconi2017heat, fodor2016far, bonilla2019active, caprini2018linear}. The absence of this
phenomenon here confirms that the effective trap potential
induced by the neighboring clusters is harmonic to a good
approximation (see also \cite{Delfau2017}).

\begin{figure}[!h]
\centering
\includegraphics[clip=true,width=0.9\columnwidth,keepaspectratio]{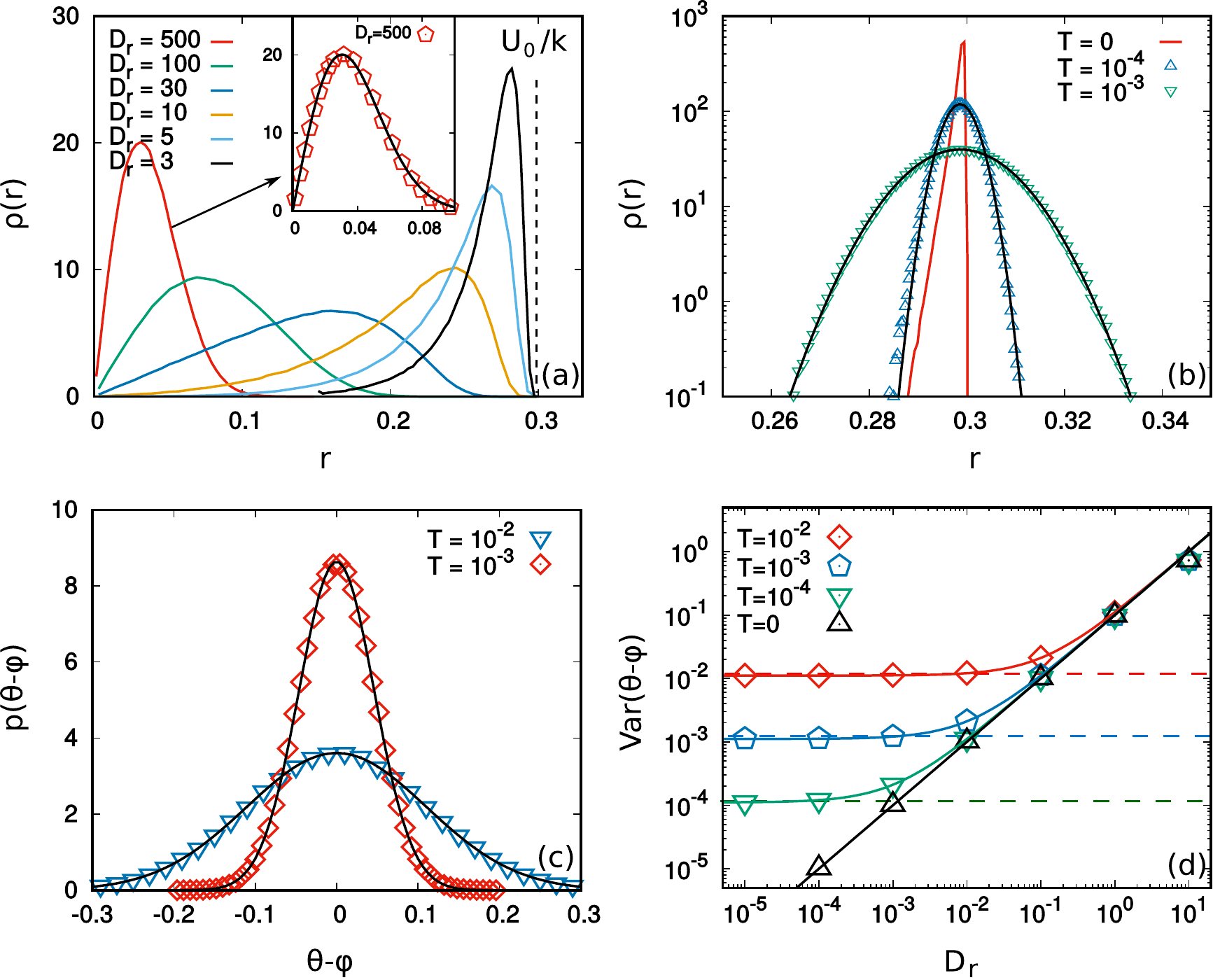}
\caption{A single ABP  particle in a harmonic trap. Panel (a): radial
probability density $\rho(r)$  for different values of
$D_r$. The inset shows a comparison between the Gaussian
approximation with some effective temperature, Eq.\eqref{eq:varianceAOUPharmonic},
and numerical data for $D_r=500$.
Panel (b): $\rho(r)$  for $D_r=0.3$ for three values of $T$: colored symbols from
numerical simulations, black lines from
Eq.\eqref{eq:radialprobT}. The red line is  for $T=0$.
Panel (c): angular probability distribution functions, $g(\theta-\phi)$, for two
different values of $T$ (colored points) compared with Eq.\eqref{eq:angularprobT} (black lines).
Panel (d): Variances of $g(\theta-\phi)$ vs $D_r$ for different values of $T$.
 Points are obtained by simulations while
continuum lines correspond to the variances predicted by the distribution \eqref{eq:radialprobT}.
The dashed lines indicate the limiting variances obtained for $D_r \to 0$.
Other parameters: $k=10$ and $U_0=3$.
}
\label{fig:harmonicprob}
\end{figure}

In the case of the ABP active force, in this effective dynamics description
in which interparticle forces are replaced by an external confining
potential, the solution of
the single-particle Fokker-Planck equation is not an easy task even
under the simple harmonic potential \cite{pototsky2012active, hennes2014self, dauchot2019dynamics, malakar2019exact}.
Recently, such a system has been studied in the presence of
hydrodynamic and steric interactions \cite{rana2019tuning} by means of the density functional theory \cite{hoell2019multi}.
In Fig.~\ref{fig:harmonicprob}(a), we numerically study the radial
density, $\rho(r)=2\pi r p(\mathbf{x})$, associated to
Eq.~\eqref{eq:xeffectivedynamics} with ABP self-propulsion. We
identify two regimes: i) when $D_r \gg k/\gamma$, the
activity plays the role of effective temperature and the
particle position density $p(\mathbf{x})$, has a Gaussian form
as in the case of AOUP and passive suspensions (see the inset of
Fig.\ref{fig:harmonicprob} (a)). In this regime, the increasing
of $D_r$ changes only the effective temperature, $\gamma
U_0^2/2D_r +  T$
( which is equivalent to $\gamma D_a + T$ discussed below Eq~\eqref{eq:varianceAOUPharmonic}), which determines the variance of the
distribution. It is straightforward to check that for large $D_r$ the variances of the ABP
and AOUP positions coincide. ii) When  $D_r \lesssim \gamma/k$,
particles arrange on a circular crown and the region near the
minimum of the harmonic potential becomes empty as in the
clusters in Fig.~\ref{fig:Snapshot}. The decrease of $D_r$
enhances the accumulation of particles in the proximity of $r^*\approx \gamma U_0/k$ (shown in Fig.~\ref{fig:harmonicprob}(a)), where
$r^*$ corresponds to the radius at which active and
the confining harmonic force balance.
 For
$T=0$ (or $T$ small enough with respect to $\gamma U_0^2/D_r$), the
distribution is strongly non-Gaussian and can be  approximated
by a Dirac $\delta$-function centered at $r-r^*$ in the limit
$D_r \to 0$. At $T=0$ the asymmetry of $\rho(r)$ is
quite evident (see Fig.~\ref{fig:harmonicprob}b)),
since as the norm of the
active force is fixed at $U_0$, particles cannot explore
regions with $r>r^*$. When $D_r \lesssim k/\gamma$,
particles arrange on the circular crown in such a way that
their orientational angles, $\theta$, can be approximated by
$\theta\approx \phi=\text{arctan}(y/x)$, i.e.
their angular coordinate with respect to the
minimum of the potential. In this case, the
probability distribution, $g(\theta-\phi)$, is a Gaussian which
becomes narrower as $D_r$ decreases (Fig.~\ref{fig:harmonicprob}c)). In
Fig.~\ref{fig:harmonicprob}(d) we study the variance of
$g(\theta-\phi)$ vs $D_r$ for several values of $T$, showing a
linear scaling (black triangles).

The effect of $T$ is shown in Fig.~\ref{fig:harmonicprob} b),
c) and d). On one hand, the larger thermal fluctuations
symmetrize the shape of $\rho(r)$ leading the system to explore the region $r> r^*$,
otherwise inaccessible. In this regime, $p(\mathbf{x})$ is
well-approximated by a Gaussian concentrated on a ring of radius
$|\mathbf{x}| \approx r^*$, so that the radial density is:
\begin{equation}
\label{eq:radialprobT}
\rho (r) \approx\mathcal{N} r \exp{\left(-\frac{k}{2T} \left( r-r^*\right)^2   \right)} \,,
\end{equation}
as confirmed in Fig.~\ref{fig:harmonicprob}(b). On the other
hand, even if $T$ does not change the Gaussian shape of
$g(\theta-\phi)$, its value can determine the variance of the
distribution. For large $D_r$ the variance is independent of
the value of $T$ (Fig.\ref{fig:harmonicprob}d)). A decrease of $D_r$ determines
a deviation from the linear behavior until a $T$-dependent
plateau is reached.
The shape of $g(\theta-\phi)$ is approximatively described by a Gaussian distribution:
\begin{equation}
\label{eq:angularprobT}
g(\theta-\phi)\approx \mathcal{N}\exp{\left(\frac{r^*}{2}\frac{\gamma U_0 }{T+\gamma D_r (r^*)^2 }\left(\phi - \theta  \right)^2  \right)} 	\,,
\end{equation}
whose variance excellently agrees with numerical simulations as shown in Fig.3 (d).
In the Supplementary Information
we show an analytical argument,
which comes from the analysis of the Fokker-Planck
equation, to derive Eqs.~\eqref{eq:radialprobT} and
\eqref{eq:angularprobT}.

The study of the distribution of particles inside each cluster,
within the present approximation of independent particles in a
confining potential allows to further understand the origin of
some differences between the AOUP and ABP active forces,
occurring at large $U_0$.
We already showed that
the constancy of the norm $U_0$ makes the difference. Another
way to see this is by noting that in the AOUP case the
fluctuations in the positions are  ruled by
$T_e/k$ given by Eq.\eqref{eq:varianceAOUPharmonic},
 which scales as
$\sim \gamma^2 D_a/\tau k^2$ for
large $\tau$. In this case, we can find a particle at $r>r^*$
with a finite probability, which is controlled by the strength
of the active force in the regime of very small $T$. A large
value of $D_a/\tau$ increases the diameter of the
cluster and the probability of finding a particle far away
from its most probable value. On the other hand, in the ABP
case fluctuations are mostly ruled by $T$, while $\gamma U_0/k$ only
determines the maximal cluster radius, without increasing the
particle positions fluctuations. Only  fluctuations induced
by $T$ could lead a particle to explore radial distances larger
than $\gamma U_0/k$ from the center of its cluster. This
difference explains why in the regime of negligible $T$, ABP
does not display the particle-exchange phase, at variance with
the AOUP model. This is also the main reason for which the AOUP
active force has a region of cluster instability for
smaller values of $U_0$ ($\sqrt{2 D_a/\tau}$). This description
directly agrees with the radial distribution measured in panel
Fig.\ref{fig:clustersize}c), where, despite the same value
of $U_0$, the probability of finding an AOUP-particle for $r
\gtrsim r^*$ is consistently larger than the ABP counterpart.
The particle-exchange phase could occur only if a ``small''
fraction of particles has an active force large enough to overcome
the effective barrier induced by the neighboring clusters,
which for $T\to 0$ can be provided only by the norm
fluctuations of $\mathbf{f}$, a mechanism which is absent in
the ABP model with $T=0$.
We remark that ``small'' has to be
considered in relation to the average number of particles
inside each cluster \cite{caprini2019transport}.

\section{Active cluster crystal in a channel}\label{sec:confining}

In this Section, we return to the original system of $N$ active
particles interacting through a repulsive GEM-$\alpha$
potential, and consider its behavior in a long
channel. The aim is studying the interplay between
cluster-crystal active aggregation and geometrical confinement.

The cluster crystal phase  obtained with soft-core interactions
and in the presence of a confining mechanism has been already
studied in \cite{wang2019melting} in the passive case. The hexagonal pattern is
stable if the cross-section of the channel is larger than the
typical interaction length of the soft-core potential, $L/R \gg
1$, a regime which will be considered hereafter.

In addition, it is well-known that active particles, also in
the non-interacting case, manifest the tendency to accumulate
near the walls  \cite{MinoClement2018EColi, li2009accumulation},
in the regime of large persistence. Both for AOUP and ABP active
forces, a microswimmer maintains its direction roughly during
the active force correlation time ($1/D_r$ or $\tau$).
 This persistence induces a profile in the space-density of the system,
 producing an anomalous maximum in front of each wall higher
 than the typical bulk density
\cite{maggi2015multidimensional, wittmann2016active, Elgeti2013wall, wensink2008aggregation, wagner2017steady}.
We, now, evaluate the relation
between the cluster crystal phase and the wall-accumulation.

\begin{figure}[!h]
\centering
\includegraphics[clip=true,width=0.9\columnwidth,keepaspectratio]{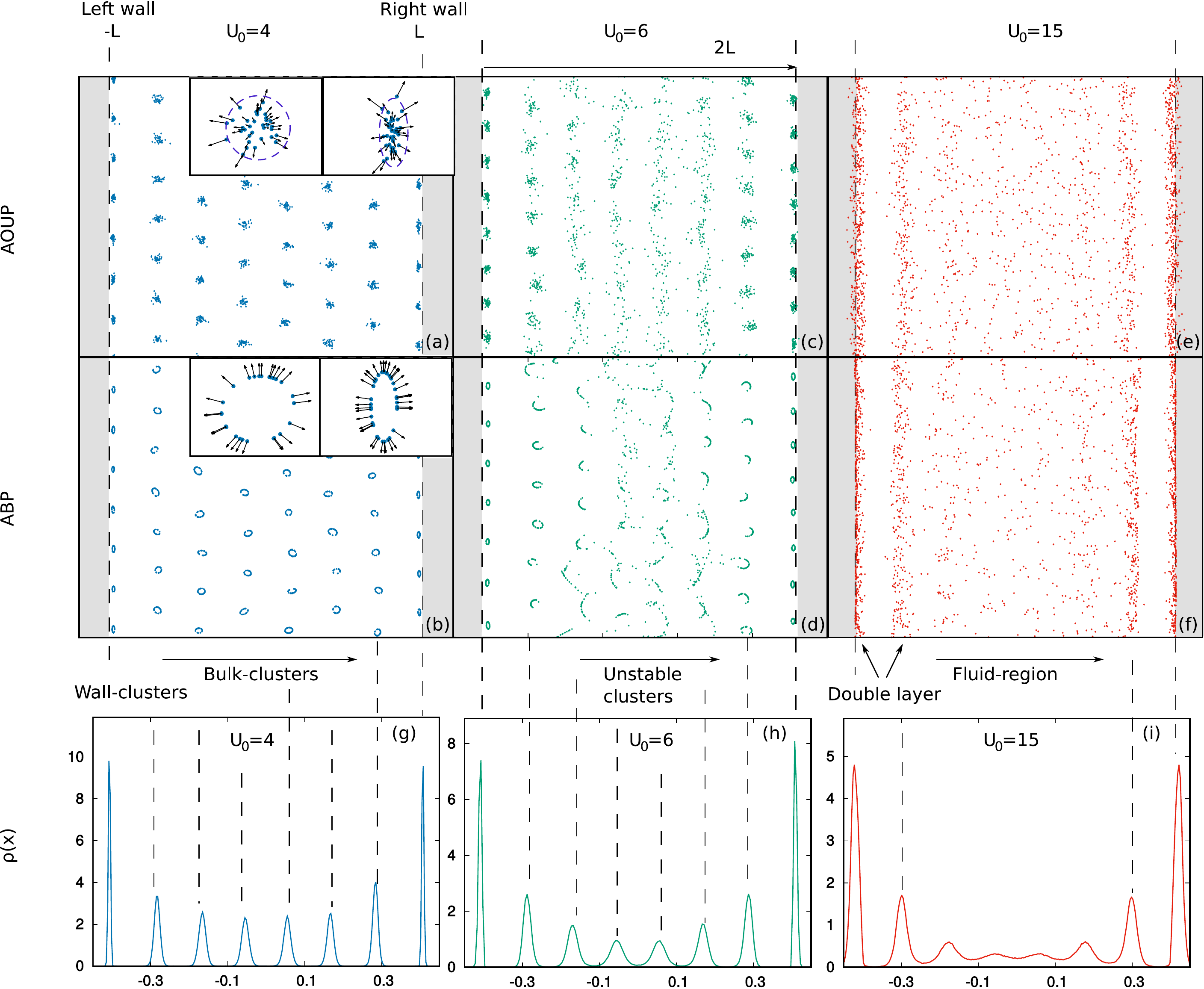}
\caption{Panels (a)-(f): Long-time snapshot for $N=2\times 10^3$
particles confined by  walls at positions $x=\pm L$. Panels (a), (c), (e) are
obtained with the AOUP model while panels (b), (d), (f) with the ABP one. From the
left to the right we increase $U_0=4,6,15$ ($\sqrt{2 D_a/\tau}$).
Grey regions are
forbidden due to the presence of the walls.
Panels (g)-(i):  spatial density for the AOUP case (similar for ABP -not shown-)
 along the transversal direction of the channel,
in correspondence of the snapshots.
Black dashed lines in correspondence of the peaks are guides to the eye.
Other  parameters are: $\epsilon=1$, $R=10^{-1}$, $L=1$, $k=10$ and $\gamma=D_r=1$.
}
\label{fig:Snapconf}
\end{figure}

We consider a channel of infinite length in the $y$-direction, and
width $2L$ in the transverse $x$-direction and model two parallel
walls by means of external repulsive potentials at positions $x=\pm
L$. For the sake of simplicity, we choose a truncated harmonic
wall-shape in such a way that the force exerted on the
particle, $\mathbf{F}_w$, is linear and directed along the
direction of the unit vector $\hat{\mathbf{x}}$. In detail:
\begin{equation}
\begin{aligned}
\mathbf{F}_w^{l,r}=-k(x \pm L) \theta(x \pm L)\hat{\mathbf{x}} \,,
\end{aligned}
\end{equation}
where the indices $l$ and $r$ refer to the left and the right
walls, respectively. $\theta(x)$ is the Heaviside function and the constant
$k$ represents the strength of the harmonic
potential. Along the $y$-direction we consider periodic
boundary conditions to mimic the infinite size of the channel.
In this way, the equations of motion read:
\begin{equation}
\label{eq:xdynamics_wall}
\gamma\dot{\mathbf{x}}_i= \bm{F}_i  + \sqrt{2 \gamma T}\,\bm{\eta}_i + \gamma\mathbf{f}_i +\mathbf{F}^l_w + \mathbf{F}^r_w \,.
\end{equation}

It is well-known that active particles show a peculiar behavior
in confined geometries \cite{wittmann2019pressure}. When the active force is strong and
persistent the microswimmers accumulate at the walls, so
 that their stationary probability distribution displays
anomalous peaks in the proximity of the walls, at $x\approx \pm
L$. The main reason is related to the time-persistence of their
motion, which keeps the direction of the active forces
roughly for a time $\sim 1/D_r$, depending on the
model employed \cite{lee2013active}. The number of particles accumulating at the
walls with respect to the particles in the bulk is controlled by
the persistence length of the active motion, $\lambda_a=U_0/D_r
\propto \sqrt{D_a \tau}$. When $\lambda_a \ll L$ the majority
of the particles moves freely in the bulk, while the number of
particles at the wall become comparable with the
bulk-particles in the opposite regime \cite{caprini2019transport}.
 In such a situation, the role of $T$ has important consequences: the accumulation
 is reduced and a bulk profile of the density occurs even far from the walls, both for ABP \cite{yan2015force, yan2018curved}
and AOUP \cite{caprini2019active} active forces.
Such a long-range effect of the wall has not a Brownian counterpart and increases with the ratio $T/\gamma  D_a$.

 In fig.~\ref{fig:Snapconf}, we display
configurations obtained for three different values of $U_0$
keeping fixed $D_r$. At small $U_0$, the cluster-crystal
 phase occurs and, for both ABP and AOUP, aligns
along the wall direction, as shown in panels a), b), c) and d).
Interestingly, we find two symmetric narrow stripes of clusters
attached to the walls whose shapes are strongly
deformed with respect to the bulk-clusters (see the insets in
panels a) and b)): in the ABP case, we observe
ellipsoidal-like clusters instead of the ring-like one, while
in the AOUP the Gaussian clusters are compressed in the
transversal direction with respect to the channel as if the
wall affects the $x$-variance of the particle distribution
inside the wall-clusters. This narrow stripe of clusters slides
along the walls changing direction with an average rate $\sim
1/D_r =\tau$, without breaking away from the walls. Clusters,
hexagonally arranged, form in the bulk at very small $U_0$,
behaving in the same way as unconfined clusters, with one of
the hexagonal directions aligned with the walls, as confirmed
by the study of the $x$-density, $\rho(x)$, in panel g) which
reveals separated peaks. This situation changes by increasing
$U_0$. In the bulk, far from the walls, clusters become
unstable: they disappear and reform continuously in time as in
the unconfined case, but maintain a clear alignment. As shown
in Fig.~\ref{fig:Snapconf} c) and d), the inner
clusters elongate along the $y$-direction until they collapse in
vertical stripes. In this case, the study of $\rho(x)$ cannot
capture this dynamical effect: the central peaks are only less
pronounced with respect to the lateral ones and overlap.
 A further increase of $U_0$ destroys the stripes order in
the inner region, creating a fluid-like homogenous bulk-phase,
whose structure is confirmed by the study of the density in
panel i) which is roughly homogenous until the occurrence of
the first stripes. In the proximity of the walls two stripes
clearly separated are stable: the one in front of the wall and
the one at the interface with the fluid-region, clearly
separated by an empty region. In practice, the wall stripe
creates an effective extra wall at distance $\sim \pm(L -R)$,
where active particles accumulate.

Finally, when $U_0$ is very large we recover the usual behavior
of active particles in the presence of walls (not shown), i.e. accumulation
near walls without any structures in the bulk region. We remark
that the suppression of any ordered structure occurs at larger
values of $U_0$ than in the case of the unconfined system. The presence
of the walls, which breaks the rotational symmetry, stabilizes
the stripe-phase which cannot form in the absence of walls.
As in the unconfined case, the drift of the pattern is
controlled by the average active force of the whole system.
Nevertheless, we observe that the confinement prevents any
motion along the transversal direction of the channel.

\section{Conclusion}

In this work, we have studied a system of active particles in
the presence of soft-core repulsive interactions,
a coarse-grained model for suspensions of
 activated complex polymers, such as dendrimers or star polymers.
We explored how the active force affects the cluster-crystal
phase of the system. We discover the existence of
traveling cluster crystals, with a speed induced by the active
force. The crystal moves coherently in space, for a
typical time which depends on the persistence of the active
force, maintaining its hexagonal structure. In addition, the
cluster-shape is deeply affected by the strength of the active
force which determines its size, until producing an unstable
region where at first clusters can exchange particles and then
destroy and reform continuously in time. Finally, for
large enough self-propulsion, the crystal melts. We have
checked in specific cases that such a phenomenology is not
restricted to our choice of interparticle potential and we
expect it to be present in a large class of soft-core
interactions. We explore two different modelizations of the
active force, both well-known in the literature, exploiting
analogies and differences between them. Besides some
differences in the particular parameter values at which
transitions occur, the only feature which distinguishes the
two descriptions is the cluster shape: for some values of the
control parameters they display a central hole in the ABP
case, but not in the AOUP. We have explained in detail the
reasons for this difference.

Finally, we confine the system into an infinitely long channel to
explore the dynamics of active soft repulsive particles. The
effective long-range effect of the wall clearly appears, deeply
influencing the structure of the pattern until to produce a
collapse into a stripe-phase aligned to the walls. Such a
phenomenon has not a passive counterpart and is entirely due to
the active force.

The consideration of more complicated confining geometries, finite-size for the particles,
and the role of repulsive and attractive potentials acting at different scales may give rise to interesting behaviors and applications of the interplay
of active forces and aggregated phases of particles.
This is material to be explored
in the close future.











\bibliography{active}


\section*{Contributions}

L.C. did the numerical simulations and most of analytical calculations. All authors contributed equally to the writing.

\section*{Additional information. Competing interests}

 E.H-G and C.L. acknowledge support from the Spanish Research Agency,
 through grant MDM-2017-0711 from the Maria de Maeztu Program for Units of Excellence in R\&D.
 C.L acknowledges useful conversations with Javier
 Aguilar-S\'anchez. The authors declare no competing financial
 interests.

\newpage
\section{Supplemental Material:}

\section*{Description of Movie 1}
The Movie is realized with $N=2\times10^3$ particles interacting with a GEM-3 potential, given by Eq.(1) of the main text with $\alpha=3$.
The Movie is composed of four panels each realized with an independent simulation with a different set of parameters.
Top panels are obtained with the AOUP model, where the self-propulsion evolves with Eq.(4) of the main text, while bottom panels are realized with the ABP active force dynamics, given by Eq.(3) of the main text.
The left column and right one are obtained by fixing $D_r=1$  (or $\tau=1$) and $D_r=0.1$ (or $\tau=10$), respectively. 
In each panel, we draw the average active force of each cluster with a green vector in the middle of each cluster.
Instead, the black arrow in the middle of the box is the average active force of the whole system.
The other parameters involved in the simulations are: $U_0=2$, $R=10^{-1}$, $L=1$, $\gamma=1$, $\epsilon=1$, $T=10^{-4} U_0^2/\gamma D_r$.

The Movie shows that both ABP and AOUP active forces give rise to a stable drifting pattern, whose typical rate of change in drift direction increases with $\tau$ (or equivalently with $1/D_r$). 

\section*{Derivation of Eqs.(12) and (13) of the main text}
\label{sec:appendixb}

To derive Eqs.(12) 
and (13) of the main text,
 it is convenient to switch from the
differential stochastic equation (2) of the main text 
 in the presence of the ABP active force, to the associated Fokker
Planck equation for the probability distribution,
$\mathcal{P}(\mathbf{x}, \theta)$, which reads:
\begin{equation}
\frac{\partial}{\partial t} \mathcal{P}=\nabla \cdot (\frac{\nabla U}{\gamma} - U_0 \bm{\hat{n}})\mathcal{P} + \frac{T}{\gamma}\nabla^2 \mathcal{P} + D_r  \frac{\partial^2}{\partial \theta^2} \mathcal{P} \,, \end{equation}
being the external potential $U$ a harmonic trap of the form:
$$
U(x,y)=\frac{k}{2}(x^2+y^2) \,.
$$
Because of the radial symmetry of $U$ we change from
 Cartesian to Polar coordinates $(x, y, \theta)
\rightarrow (r, \phi, \theta)$, being $r=\sqrt{x^2+y^2}$ and
$\phi=\arctan{\left(y/x\right)}$, in such a way that the
Fokker-Planck equation for the probability distribution
function $\tilde{\mathcal{P}}(r, \phi, \theta)$ becomes:
\begin{equation}
\label{eq:ap_FPpolar}
\frac{\partial}{\partial t}\tilde{\mathcal{P}} =
\frac{\partial}{\partial r} \left[\frac{k}{\gamma} r \tilde{\mathcal{P}} - \frac{T}{\gamma r} \tilde{\mathcal{P}} +
\frac{T}{\gamma}\frac{\partial}{\partial r}\tilde{\mathcal{P}}  -U_0 \tilde{\mathcal{P}} \,\cos{(\theta-\phi)}\right] +
\left[\frac{T}{\gamma r^2}\frac{\partial^2}{\partial \phi^2}\tilde{\mathcal{P}} +\frac{U_0}{r}\tilde{\mathcal{P}} - \frac{U_0}{r}\sin{(\theta-\phi)}\frac{\partial}{\partial \phi}\tilde{\mathcal{P}} \right] + D_r \frac{\partial^2}{\partial \theta^2}\tilde{\mathcal{P}} \,.
\end{equation}
Finding an exact solution of such a partial differential
equation is not so easy. 
Assuming that $\theta \sim \phi$, we can approximate $\sin{\left( \phi
-\theta \right)}\approx \phi- \theta$ and $\cos{\left(\phi -
\theta  \right)} \approx 1$, neglecting terms of order $(\phi-\theta)^2$ and higher. 
This approximation can hold only in the limit $D_r \to 0$.
Assuming that the radial flux (i.e. the first square brackets in Eq.\eqref{eq:ap_FPpolar}) is zero, we can find an
approximate stationary solution for the radial component of $\tilde{\mathcal{P}}$,
under the assumption that $\tilde{\mathcal{P}}$ is factorized as the product of a radial component, $\rho(r)$, and an angular component, $f(\theta, \phi)$. 
The vanishing of the
first square bracket in Eq. (\ref{eq:ap_FPpolar}) leads to
\begin{flalign}
 \tilde{\mathcal{P}}(r, \theta, \phi) \approx f(\theta, \phi)\,\, r\, \,
 \exp{\left(\frac{\gamma}{T}\left[-\frac{k}{\gamma}\frac{r^2}{2} +U_0 r\right]  \right)}\, ,
\end{flalign}
where $f(\theta, \phi)$ is to be
determined. 
The radial part of this expression is proportional to $\rho(r)$, 
and
thus $p(\mathbf{x})$ (switching to Cartesian components) is a Gaussian
with variance $T/k$ centered around a ring of radius $r^*=\gamma U_0/k$.

Since circular symmetry of the clusters implies translational invariance of the angular components, it is straighforward to derive that $f(\theta, \phi)=g(\phi-\theta)$.
Approximating $r \approx r^*$ in the remaining terms of 
Eq.~\eqref{eq:ap_FPpolar} and using that $\sin{\left( \phi -\theta
\right)}\approx \phi- \theta$, 
we find 
\begin{flalign}
\label{eq:appendix_approx_thetaphiprob}
 g(\phi-\theta) \approx \mathcal{N}  \exp{\left(-\frac{\gamma U_0 r^*}{T+\gamma D_r(r^*)^2}\frac{\left[\phi -\theta \right]^2}{2}  \right)} \, ,
\end{flalign}
with  $\mathcal{N}$ a normalization factor. 
Eq.\eqref{eq:appendix_approx_thetaphiprob}
shows that $\phi$ is distributed as a Gaussian 
centered at $\theta$,
whose variance is
$$\text{Var}(\phi-\theta)=\frac{T+\gamma D_r (r^*)^2}{\gamma U_0 r^*} \,,$$
and demonstrates Eq.(13) of the main text.
Combining both results we get an approximate solution of the Fokker-Planck equation, whose validity is restricted to the limit $D_r \to 0$.







\end{document}